\documentclass[twocolumn,preprintnumbers,amsmath,amssymb]{revtex4}
\usepackage{graphicx}
\usepackage{float}
\usepackage[usenames,dvipsnames]{color}
\begin{document}
\title{Temperature Dependence of the Neutron Lifespan
}
\author{Cheng Tao Yang${}^a$, Jeremiah Birrell${}^b$, Johann Rafelski${}^a$}
\affiliation{${}^a$Department of Physics, The University of Arizona, Tucson, Arizona 85721, USA\\
${}^b$Department of Mathematics and Statistics, University of Massachusetts Amherst, Amherst, MA 01003, USA}
\date{\today}


\begin{abstract}

{Current precision big bang nucleosynthesis (BBN) studies} motivate us to revisit the neutron lifespan in the plasma medium of the early universe. The mechanism we explore is the Fermi-blocking of decay electrons and neutrinos by plasma. As result, neutrons live longer and we find a significant 6.4\% modification of neutron abundance in the BBN era arising from in plasma modification of the neutron lifetime. This effect can influence the final abundances of the light elements in BBN.

\end{abstract}
\maketitle


\section{Introduction}\label{Intro}
Element production during BBN is influenced by several parameters, e.g. baryon to photon ratio $\eta_b$, number of neutrino species $N_\nu$, and neutron to proton ratio, as controlled by both the dynamics of neutron freeze-out at temperature $T_f\approx 0.8\mathrm{MeV}$ and neutron lifetime. Since about 200 seconds pass between neutron freeze-out and BBN at $T\approx0.07\mathrm{MeV}$, the neutron lifetime is one of the important parameter controlling BBN element yields~\cite{Pitrou:2018cgg}. However, the neutron decay in cosmic plasma medium has seemingly never been considered before. Therefore, here we study the Fermi suppression effects which lengthen the neutron lifetime in the early universe. The medium dependence of particle decay was recognized by Kuznetsova et al~\cite{Kuznetsova:2010pi}, we will use their method to study the plasma effects on neutron decay in the cosmic medium.

After freeze-out, the neutron abundance is subject to natural decay
\begin{align}\label{Ndec}
n\longrightarrow p+e+\overline{\nu}_e\;,
\end{align}
where the current experimental neutron lifetime quoted by the Particle Data Group~\cite{Patrignani:2016xqp} is $\tau_n^0=880.2\pm1.0\,\mathrm{sec}$. The latest measurement for neutron lifetime by using magnetogravitational trap is $877.7\pm0.7\,\mathrm{sec}$ \cite{Pattie:2018vsj}. In the standard Big Bang Nucleosynthesis, the neutron abundance when nucleosynthesis begins is given by~\cite{Pitrou:2018cgg}
\begin{align}
\label{Xn_abundance}
X_n(T_{BBN})=X_n^f\exp\left(-\frac{t_{BBN}-t_f}{\tau_n^0}\right)\approx0.13\;,
\end{align}
The normalizing neutron freeze-out yield $X_n^f$ 
\begin{align}
\label{Xn_abundance2}
X_n^f \equiv  \frac{n_n^f}{n_n^f+n_p^f}= \frac{n_n^f/n_p^f}{1+n_n^f/n_p^f}\;.
\end{align}
where $n_n^f$ and $n_p^f$ are freeze-out neutron and proton densities, respectively. The thermal equilibrium yield ratio is
\begin{align}
\label{Xn_abundance3}
 \frac{n_n^f}{n_p^f}= \exp\left(-Q/T_f\right)\;,\qquad Q=m_n-m_p\;,
\end{align}
assuming a instantanous freeze-out, depends on temperature $T_f$ at which neutrons decouple from the heat bath, and the neutron-proton mass difference (in medium). The values considered  are in the range $X_n^f=0.15\sim0.17$ \cite{Pitrou:2018cgg}. A dynamical approach to neutron freezeout is necessary to fully understand $X_n^f$, we hope to return to this challenge in the near  future.

Following freeze-out the neutron is subject to natural decay and normally the neutron lifetime in vacuum $\tau_n^0$ is used c.f. Eq.\,(\ref{Xn_abundance}) to calculate the neutron abundance resulting in the \lq desired\rq\ value $X_n(T_{BBN})\approx0.13$ when BBN starts. To improve precision a dynamically evolving neutron yield needs to be studied and for this purpose we explore here the neutron decay which occurs in  medium, not vacuum. This leads to  neutron lifespan dependence on temperature of the cosmic medium as the decay occurs for a particle emerged in plasma of electron/positron, neutrino/antineutrino, (and protons).

Two physical effects of the medium  influence the neutron lifetime in the early universe noticeably:
\begin{itemize}
\item Fermi suppression factors from the medium: 
During the temperature range $T_f\geqslant T\geqslant T_{BBN}$, electrons and neutrinos in the background plasma can reduce the neutron decay rate by Fermi suppression to the neutron decay rate. Furthermore, the neutrino background can still provide the suppression after electron/positron pair annihilation becomes nearly complete.
\item Photon reheating:
When $T\ll m_e$ the electron/positron annihilation occurs, the entropy from $e^\pm$ is fed into photons, leading to photon reheating. The already decoupled (frozen-out) neutrinos remain undisturbed. Therefore, after annihilation we have two different temperatures in cosmic plasma: neutrino temperature $T_\nu$ and the photon and proton temperature $T$ respectively.
\end{itemize}
These two effect will be included in the following exploration of the neutron lifetime in the early universe as a function of $T$. We show how these effects alter the neutron lifespan and obtain the modification of the neutron yield at the time of BBN. Yet another effect was considered by Kuznetsova et al~\cite{Kuznetsova:2010pi} which is due to time dilation originating in particle thermal motion. In our case for neutrons with $T/m<10^{-3}$ this effect is negligible. Below we will explicitly assume that the neutron decay is studied in the neutron rest frame.

In section~\ref{Rate_Medium} we discuss the general form of the neutron decay rate in a heat bath, including the Fermi suppression factors from the medium. In section~\ref{Reheating} we derive the relation between neutrino temperature and photon temperature in early universe. In section~\ref{Neutron} we compute the modified neutron abundance due to effects of the medium and compare with the standard (vacuum) result. The implications of these results will be discussed in section~\ref{Disscusion}.


\section{Decay Rate in Medium}\label{Rate_Medium}

The invariant matrix element for the neutron decay Eq.\,(\ref{Ndec}) for non-relativistic neutron and proton is given by
\begin{align}
\langle|\mathcal{M}|^2\rangle\approx16\,G^2_FV^2_{ud}\,m_nm_p(1+3g^2_A)(1+RC)E_{\bar{\nu}}E_e,
\end{align}
where the Fermi constant is $G_F=1.1663787\times10^{-5}\,\mathrm{GeV}^{-2}$, $V_{ud}=0.97420$ is an element of the Cabibbo-Kobayashi-Maskawa (CKM) matrix~\cite{Czarnecki:2018okw,Marciano:2005ec,Czarnecki:2004cw}, and $g_A=1.2755$ is the axial current constant for the nucleons~\cite{Czarnecki:2018okw,Marciano:2014ria}. We also consider the total effect of all radiative corrections relative to muon decay that have not been absorbed into Fermi constant $G_F$. The most precise calculation of this correction~\cite{Marciano:2014ria,Marciano:2005ec} gives $(1+RC)=1.03886$. 

In the early universe the neutron decay rate in medium, at finite temperature can be written as~\cite{Kuznetsova:2010pi}
\begin{align}
\frac{1}{\tau^\prime_n}=\frac{1}{2m_n}\int&\frac{d^3p_{\bar{\nu}}}{(2\pi)^32E_{\bar{\nu}}}\frac{d^3p_p}{(2\pi)^32E_p}\frac{d^3p_e}{(2\pi)^32E_e}\notag\\
&(2\pi)^4\delta^4\left(p_n-p_p-p_e-p_{\bar{\nu}}\right)\langle|\mathcal{M}|^2\rangle\notag\\
&\big[1-f_p(p_p)\big]\big[1-f_e(p_e)\big]\big[1-f_{\bar{\nu}}(p_{\bar{\nu}})\big]\;,
\end{align}
where we consider this expression in the rest rest frame of neutron, {\it i.e.\/} $p_n=(m_n,0)$. The phase-space factors $(1-f_i)$ are Fermi suppression factors in the medium. The Fermi-Dirac distributions for electron and non-relativistic proton are given by
\begin{align}
&f_e=\frac{1}{e^{E_e/T}+1},\\
&f_p=e^{-E_p/T}=e^{-m_p/T}\,e^{-p_p^2/2m_pT}.
\end{align}
For neutrinos, after neutrino/antineutrino kinetic freeze-out they become free streaming particles. If we assume that kinetic freeze out occurs at some time $t_k$ and temperature $T_k$, then for $t>t_k$ the free streaming distribution function can be written as~\cite{Birrell:2012gg}
\begin{align}
f_{\bar{\nu}}=\frac{1}{\exp{\left(\sqrt{\frac{E^2-m_\nu^2}{T_\nu^2}+\frac{m^2_\nu}{T^2_k}}+\frac{\mu_{\bar{\nu}}}{T_k}\right)+1}},
\end{align}
for antineutrinos and we define the effective neutrino temperature $T_\nu$ as
\begin{align}
T_\nu\equiv\frac{a(t_k)}{a(t)}T_k.
\end{align}
In the following calculation, we assume the condition $T_k\gg\mu_{\bar{\nu}},\,m_\nu$, {\it i.e.\/} we consider the massless neutrino in plasma. Substituting the distributions into the decay rate formula and using the neutron rest frame, the decay rate can be written as 
\begin{align}
\label{Decay_rate_01}
\frac{1}{\tau_n^\prime}&=\frac{G^2_FQ^5V^2_{ud}}{2\pi^3}\,(1+3g^2_A)\,(1+RC)\\
&\times\int^1_{m_e/Q}d\xi\,\frac{\xi(1-\xi)^2}{\exp\left(-Q\xi/{T}\right)+1}\frac{\sqrt{\xi^2-(m_e/Q)^2}}{\exp\left(-Q(1-\xi)/T_\nu\right)+1},\notag
\end{align} 
where $Q$ was defined in Eq.\;(\ref{Xn_abundance3}) and we integrate using dimensionless variable $\xi=E_e/Q$. From Eq.(\ref{Decay_rate_01}), the decay rate in vacuum can be written as
\begin{align}
&\frac{1}{\tau_n^0}=\frac{G^2_Fm_e^5V^2_{ud}}{2\pi^3}(1+3g^2_A)\,(1+RC)\,f^\prime,
\end{align}
where the phase space factor $f^\prime$ is given by
\begin{align}
f^\prime&\equiv\left(\frac{Q}{m_e}\right)^5\int^1_{m_e/Q}d\xi\,{\xi(1-\xi)^2}\sqrt{\xi^2-(m_e/Q)^2}
\notag\\&=1.6360.
\end{align}
The phase space factor is also modified by the Coulomb correction between electron and proton, proton recoil, nucleon size correction etc. This has been studied by Wilkinson~\cite{Wilkinson:1982hu}, and the phase space factor is given by~\cite{Czarnecki:2018okw, Czarnecki:2004cw,Wilkinson:1982hu}
\begin{align}
f=1.6887.
\end{align}
These effect amount to adding the factor $\mathcal{F}$ to our calculation
\begin{align}
\mathcal{F}=\frac{f}{f^\prime}=1.0322,
\end{align}
then the neutron lifespan can be written as
\begin{align}
\tau^{\mathrm{Vacuum}}_n=\frac{\tau^0_n}{\mathcal{F}}=879.481\,\mathrm{sec},
\end{align}
which compare well to the experiment result $877.7\pm0.7\,\mathrm{sec}$ \cite{Pattie:2018vsj}. In the case of plasma medium, we do not expect that these effect (Coulomb correction between electron and proton, proton recoil, nucleon size correction etc) are modified in the cosmic plasma. Thus we adapt the factor into our calculation and the neutron decay rate in the cosmic plasma is given by
\begin{align}
\label{Decay_rate_02}
&\frac{1}{\tau_n^{\mathrm{Medium}}}=\frac{G^2_FQ^5V^2_{ud}}{2\pi^3}\,(1+3g^2_A)\,(1+RC)\,\mathcal{F}\\
&\times\int^1_{m_e/Q}d\xi\,\frac{\xi(1-\xi)^2}{\exp\left(-Q\xi/{T}\right)+1}\frac{\sqrt{\xi^2-(m_e/Q)^2}}{\exp\left(-Q(1-\xi)/T_\nu\right)+1}.\notag
\end{align}
From Eq.(\ref{Decay_rate_02}) we see that the neuron decay rate in the early universe depends on both the photon temperature $T$ and the neutrino effective temperature $T_\nu$.


\section{Photon Reheating}\label{Reheating}

\begin{figure}[t]
\begin{center}
\includegraphics[width=3.5in]{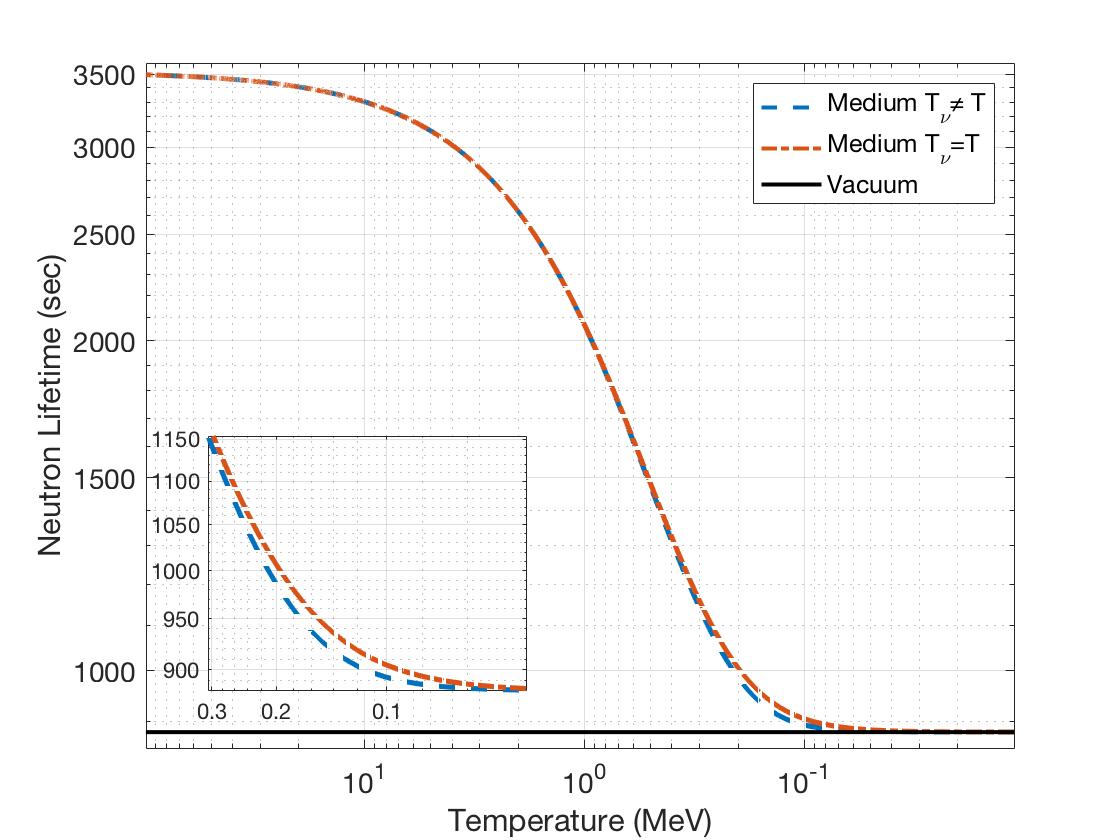}
\caption{The neutron lifetime $\tau_n^\mathrm{Medium}$ in the cosmic plasma as a function of temperature. At high temperature $T=100\,\mathrm{MeV}$ the neutron lifetime is $3495\,\mathrm{sec}$ which is $3.974$ times larger than the lifetime in vacuum. At low temperature, $T<m_e$, the neutron lifetime depends also on the neutrino temperature, $T_\nu$, the effect is amplified in the insert.}
\label{Decay_Rate}
\end{center}
\end{figure}

After neutrino free-out and when $m_e\gg T$, the $e^{\pm}$ becomes non-relativistic and annihilate. In this case, their entropy is transferred to the other relativistic particles still present in the cosmic plasma, {\it i.e.\/} photons, resulting in an increase in photon temperature as compared to the freestreaming neutrinos. From entropy conservation we have
\begin{align}
\label{Entropy}
\frac{2\pi}{45}g^s_\ast(T_k)T^3_kV_k+S_{\nu}(T_k)=\frac{2\pi}{45}g^s_\ast(T)T^3V+S_{\nu}(T),
\end{align}
where we use the subscripts $k$ to denote quantities for neutrino freezeout and $g^s_\ast$ counts the degree of freedom for relativistic species in early universe. After neutrino freezeout, their entropy is conserved independently and the temperature can be written as
\begin{align}
T_\nu\equiv\frac{a(t_k)}{a(t)}T_k=\left(\frac{V_k}{V}\right)^{1/3}T_k.
\end{align}
In this case, from entropy conservation, Eq.(\ref{Entropy}), we obtain
\begin{align}
\label{Neutrino_Photon}
T_\nu=\frac{T}{\kappa},\,\,\,\,\,\,\kappa\equiv\left[\frac{g^s_\ast(T_k)}{g^s_\ast(T)}\right]^{1/3}.
\end{align}
From Eq.(\ref{Neutrino_Photon}) the neutron decay rate in a heat bath can be written as
\begin{align}
\label{Decay_rate_03}
&\frac{1}{\tau_n^\mathrm{Medium}}= \frac{G^2_FQ^5V^2_{ud}}{2\pi^3}(1+3g^2_A)\,(1+RC)\,\mathcal{F}\\
&\times\int^1_{m_e/Q}d\xi\,\frac{\xi(1-\xi)^2}{\exp\left(-Q\xi/{T}\right)+1}\frac{\sqrt{\xi^2-(m_e/Q)^2}}{\exp\left(-Q(1-\xi)\kappa/T\right)+1}.\notag
\end{align}
In the high temperature regime, $T\gg Q$, the exponential terms in the Fermi distribution becomes $1$ and the decay rate is given by
\begin{align}
&\frac{1}{\tau_n^\mathrm{Medium}}=\frac{1}{4}\left(\frac{1}{\tau_n^\mathrm{Vacuum}}\right)\;,
\qquad
T\gg Q\;.
\end{align}
In Fig(\ref{Decay_Rate}), we plot the the neutron lifetime $\tau^\mathrm{Medium}_n$ in plasma as a function of temperature. Fermi-suppression from electron and neutrino increases the neutron lifetime as compared to value in vacuum. At low temperature, $T<m_e$, most of the electrons and positrons have annihilated and the main Fermi-blocking comes from the cosmic neutrino background. In this regime, the neutron lifetime depends also on the neutrino temperature, $T_\nu$. For cold neutrinos $T_\nu<T$, the Fermi suppression is smaller than the hot one $T_\nu=T$. 
\begin{figure}[h]
\begin{center}
\includegraphics[width=3.5in]{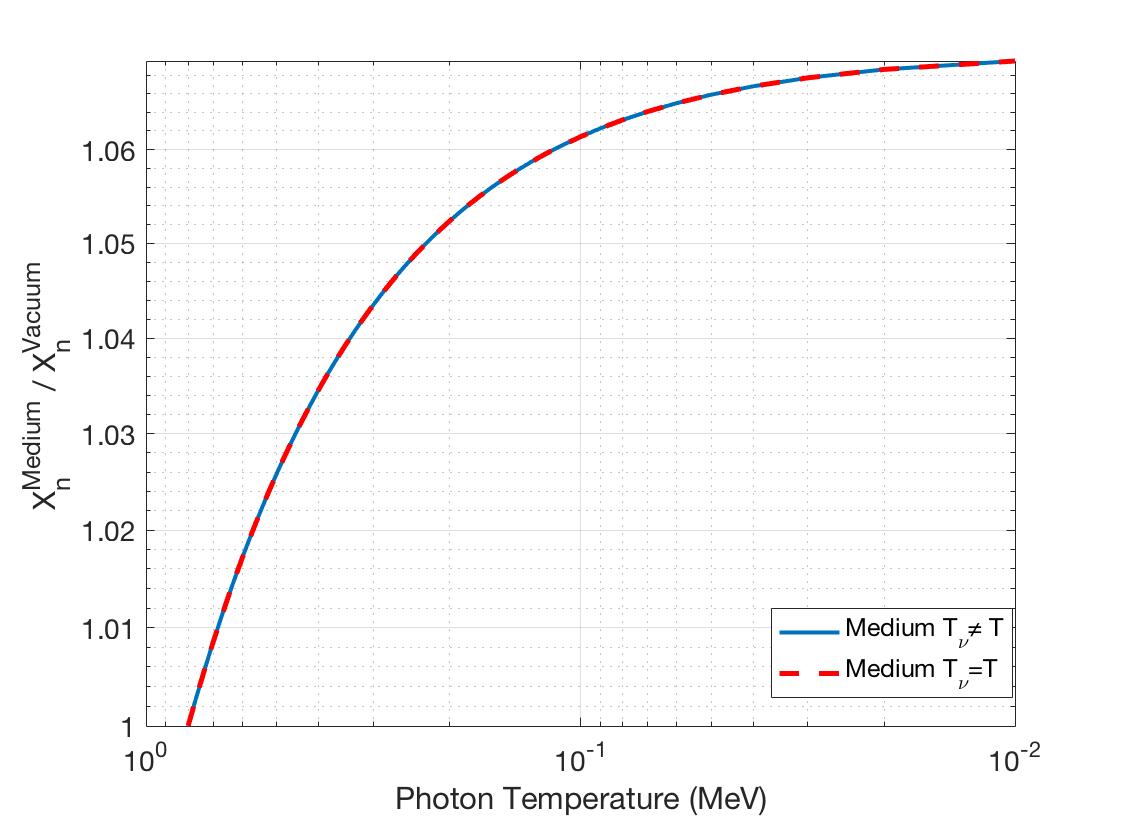}
\caption{The neutron abundance ratio as a function of temperature. Consider the neutron freezeout temperature $T_f=0.08\mathrm{MeV}$ and the BBN temperature $T_{BBN}\approx0.07\mathrm{MeV}$, we find the abundance ratio ${X_n^{\mathrm{Meduim}}}/{X_n^{\mathrm{vacuum}}}=1.064$ at temperature $T_{BBN}$.}
\label{Neutron_Abundance}
\end{center}
\end{figure}

\section{Neutron Abundance}\label{Neutron}

After the neutron freezeout, the neutron abundance is only affected by the neutron decay. The neutron concentration can be written as
\begin{align}
\label{Abundance}
X_n=X_n^f\,\exp\bigg[-\int^t_{t_f}\frac{dt^\prime}{\tau_n}\bigg],
\end{align}
where we use the subscripts $f$ to denote quantities at neutron freezeout. Using Eq.(\ref{Decay_rate_03}) and Eq.(\ref{Abundance}), the neutron abundance ratio between plasma medium and vacuum is given by
\begin{align}
\label{Abundance_Ratio}
\frac{X_n^{\mathrm{Meduim}}}{X_n^{\mathrm{Vacuum}}}=\exp\bigg[-\int^t_{t_f}dt^\prime\left(\frac{1}{\tau^\prime_n}-\frac{1}{\tau^0_n}\right)\bigg].
\end{align}
In Fig.(\ref{Neutron_Abundance}), we plot the neutron abundance ratio as a function of temperature. Consider the neutron freezeout temperature $T_f=0.08\mathrm{MeV}$ and the BBN temperature $T_{BBN}\approx0.07\mathrm{MeV}$, we found that the ratio ${X_n^{\mathrm{Meduim}}}/{X_n^{\mathrm{Vacuum}}}=1.064$ at temperature $T_{BBN}$. Then from Eq.(\ref{Xn_abundance}) the neutron abundance in plasma medium is given by
\begin{align}
X_n^{\mathrm{Meduim}}=1.064\,X_n^{\mathrm{Vacuum}}\approx0.138.
\end{align}
In this case, the neutron abundance will increase about $6.4\%$ in the cosmic plasma which should affect the final abundances of the Helium-4 and other light elements in BBN.


\section{Conclusion and Discussion}\label{Disscusion}

One of the important parameters of standard BBN is the neutron lifetime, as it affects the neutron abundance after neutron freezeout at temperature $T_f\approx 0.8 \mathrm{MeV}$ and before the BBN $T\approx0.07 \mathrm{MeV}$. In the standard BBN model, it is necessary to have a neutron-to-proton ratio $n/p\approx1/7$ when BBN begins in order to explain the observed values of hydrogen and helium abundance~\cite{Pitrou:2018cgg}.

In this paper we have evaluated the effect of Fermi suppression on the neutron lifetime due to the background electron and neutrino plasma. We found that in medium the neutron lifetime is lengthened by upt to a factor 4 at a high temperature ($T>10$\,MeV). Our method should in principle also be considered in the study of medium modification of just about any of the BBN weak interaction rates, this remains a task for another day.

In the temperature range between neutron freeze-out just below $T=1$\;MeV and BBN conditions the effect of neutron lifespan is smaller but still noticeable. Near neutron freeze-out both decay electron and neutrino are blocked. However, after $e^\pm$ annihilation is nearly complete closer to BBN Fermi-blocking comes predominantly from the cosmic neutrino background and the neutron lifetime depends on the temperature $T_\nu<T$.

We found that the increased neutron lifetime results in an increased neutron abundance of ${X_n^{\mathrm{Meduim}}}/{X_n^{\mathrm{vacuum}}}=1.064$ at $T_{BBN}\approx0.07 \mathrm{MeV}$ {\it i.e.\/} we find a $6.4\%$ \emph{increase} in neutron abundance due to the medium effect at the time of BBN. We believe that this effect needs to be accounted for in the precision study of the final abundances of hydrogen, helium and other light elements produced in BBN.


\end{document}